\documentstyle[aps,epsf,twocolumn]{revtex}   
\begin{document}   
 
\title{Repulsion of Resonance States and Exceptional Points } 
  
\author{W.D. Heiss}

\address{Centre for Nonlinear Studies and
Department of Physics\\
 University of the Witwatersrand,
PO Wits 2050, Johannesburg, South Africa}

\maketitle


\begin{abstract}   
Level repulsion is associated with exceptional points which are square root
singularities of the energies as functions of a (complex) interaction
parameter. This is also valid for resonance state energies. Using this concept
it is argued that level anti-crossing (crossing) must imply crossing 
(anti-crossing) of the corresponding widths of the resonance states. 
Further, it
is shown that an encircling of an exceptional point induces a phase change
of one wave function but not of the other. An experimental setup is discussed 
where this phase behaviour which differs from the one encountered at a diabolic
point can be observed.
\end{abstract}    

\vskip 1cm

The dependence on parameters of the energies and widths of resonance 
states has always been a central focus of interest in
virtually all domains of physics. One particular aspect is the repulsion 
of levels in the complex energy plane. A level repulsion in the 
complex energy plane can appear as a crossing of, say, their 
real parts,  since the
corresponding imaginary parts still can avoid each other; likewise, a genuine 
repulsion of the real parts can imply a crossing of the imaginary parts. These 
aspects have been discussed in a variety of contexts: in nuclear and particle
physics, for electro-magnetic resonators \cite{bren,rot} and in results found 
for absorptive media in solid state 
physics \cite{ever}. An investigation on a more theoretical footing is found 
in \cite{mond}.

The purpose of the present paper is twofold. First, we demonstrate that the 
various types of crossing and/or anti-crossing can be understood from a common
principle. They are related to each other by the position 
of particular singular points of the spectrum, which are called 
exceptional points (EP) \cite{kato}. The
second aspect deals with the fact that, if an EP is encircled, the phases
of the associated wave functions change in a particular way which is different
from the phase behaviour when a genuine degeneracy of levels (a diabolic point)
is encircled \cite{ber}. At an EP two levels coalesce, but,
as is discussed below, an EP is not to be confused with a genuine degeneracy
of two resonant states. The fact that there are different types of coalescence
of resonance states was pointed out in \cite{mond}. However,
in the quoted paper, the type of singularity, in fact the concept of an 
EP was not explicitly employed; rather the effect upon the Green's function
or the scattering amplitude was elaborated, which is of lesser interest here. 
Genuine degeneracies of resonance states have been discussed 
in the literature, 
including an associated phase behaviour of the wave functions involved 
\cite{nenc,mim}. However, the subject of the present paper which is a
generalisation and further expansion of a previous publication \cite{hei}
addresses a thoroughly different situation.

All essential aspects of exceptional points can be illustrated on an 
elementary level with a two level model. In fact, for finite or infinite
dimensional problems an isolated exceptional point can be described locally
by a two dimensional problem \cite{hest}. In other words, even though a
high or infinite dimensional problem is globally more complex than the
two dimensional problem, we do not loose generality for our specific purpose 
when the restriction to a two dimensional problem is made. For easy
illustration we begin with the discussion of
\begin{equation}  \label{ham}
H=\pmatrix {\epsilon _1 & 0 \cr 0 & \epsilon _2}+
\lambda U \pmatrix {\omega _1 & 0 \cr 0 & \omega _2} U^{\dagger}
\end{equation}
with
\begin{equation} \label{uang}
U(\phi )=\pmatrix {\cos \phi & -\sin \phi \cr \sin \phi & \cos \phi }.
\end{equation}
This is, up to a similarity transformation, the most general form of a real
two dimensional Hamilton matrix of the type $H_0+\lambda H_1$. We emphasize 
again that our aim is not in
particular directed at a physical model that is describable by a two
dimensional problem although there may exist interesting problems in our
special context. The example has been chosen for illustration, while
the physical application that we have in mind is in general an infinite 
dimensional situation.

The eigenvalues of $H$ are given by
\begin{equation}  \label{eigv}
E_{1,2}(\lambda )={\epsilon _1+\epsilon _2+\lambda (\omega _1+\omega _2)
\over 2} \pm R
\end {equation}
where
\begin{eqnarray} \label{res}
R&=&\biggl\{({\epsilon _1-\epsilon _2\over 2})^2  \\
&+&({\lambda (\omega _1-\omega _2)\over 2})^2+{1\over 2}
\lambda (\epsilon _1-\epsilon _2)(\omega _1-\omega _2)\cos 2 \phi
\biggr\}^{1/2}. \nonumber
\end{eqnarray}
Clearly, when $\phi =0$ the spectrum is given by the two lines
$$E_k^0(\lambda )=\epsilon _k+\lambda \omega _k, \quad k=1,2 $$
which intersect at the point of degeneracy
$\lambda =-(\epsilon _1-\epsilon _2)/(\omega _1-\omega _2)$. When the
coupling between the two levels is turned on by switching on $\phi $, the
degeneracy is lifted and an avoided level crossing occurs. Now the two levels 
coalesce in the complex $\lambda $-plane where 
$R$ vanishes. This happens at the complex conjugate points
\begin{equation} \label{exc}
\lambda _c=-{\epsilon _1-\epsilon _2\over \omega _1-\omega _2}
\exp (\pm 2i\phi ).
\end{equation}
At these points, the two levels $E_k(\lambda )$ are connected by a square
root branch point; in fact the two levels are the values of one analytic
function on two different Riemann sheets. Obviously, this connection is not 
of the type encountered at a genuine diabolic point. We stress again that the 
same nature of singularity prevails also in an $N-$dimensional matrix problem 
of the type $H_0+\lambda H_1$.

\begin{figure}
\epsfxsize=3.0in
\centerline{
\epsffile{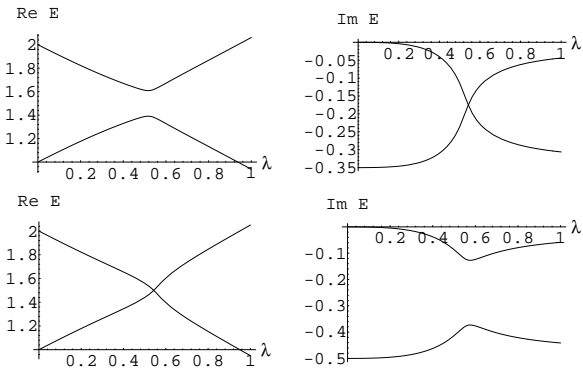}}
\vglue 0.15cm
\caption{
Level repulsion (width crossing) and level crossing (width repulsion) for
smaller (top, $\mu =0.35$) and larger (bottom, $\mu =0.5$) absorption 
$\mu $ which is chosen such that the EP lies
just below the real $\lambda -$axis in the former and just above
in the latter case. The other parameters are $\epsilon _1=1,
\epsilon _2=2,\omega _1=1,\omega _2=-1,\phi _1=0.2$.
}
\label{fig1}
\end{figure}

The question arises as to whether the existence of the EPs is of physical 
interest in addition to their pertinent association with level repulsion.
Before we turn to actual proposals of encircling an EP in an experiment
we first discuss formally the various effects of encircling an EP. Obviously, 
we we obtain the same information by comparing the results of two different 
sweeps over appropriate values of $\lambda $, the one by passing 
an EP on its left hand side and the other 
on its right hand side. In principle, this 
can be achieved by choosing complex values of $\lambda $. In order to get
closer to an actual experimental situation we expand the Hamiltonian by 
considering the enlarged model
\begin{eqnarray} H=
\pmatrix{\epsilon _1 & 0 \cr 0 & \epsilon _2}+\lambda 
U(\phi _1)\pmatrix{ \omega _1 & 0 \cr 0 & \omega _2}U^{\dagger }(\phi _1)
\nonumber  \\  \label{hcomp}
- i \mu
U(\phi _2)\pmatrix{\sigma _1 & 0 \cr 0 & \sigma _2}U^{\dagger }(\phi _2).
\end{eqnarray}
The additional term ($\mu $ real) can be used to describe an absorption 
while adhering to real values of $\lambda $. Also, the unperturbed energies
$\epsilon _k$ may be chosen complex, that is including a width.
The EP of the enlarged model are situated at
\begin{equation} \label{ep}
\lambda _c=(-1+i\mu e^{\pm 2i\phi _2}{\sigma _1-\sigma _2\over
\epsilon _1-\epsilon_2})\cdot e^{\pm 2i\phi _1} {\epsilon _1-\epsilon_2
\over \omega _1-\omega _2}.
\end{equation}

In an experimental situation a judicious choice of these additional 
parameters can move one of the EP in the $\lambda -$plane close to the real 
axis. In fact, we now demonstrate that the position of an EP can be arranged 
in various ways to lie just above or below the real $\lambda -$axis. The
different effect of these two situations for
the energies, when sweeping over real values of $\lambda $, turns out to be
anti-crossing for the real and crossing for the imaginary parts in the one 
case, and crossing for the real and anti-crossing for the imaginary parts in
the other.

In Fig.1 we illustrate the real and imaginary parts of the energies as a 
function of the real parameter $\lambda $ for two different values of $\mu $.
For simplicity, the choice $\phi _2=0, \sigma _1=1$ and $\sigma _2=0$ was made.
Other choices lead qualitatively to the same result if $\sigma _1\ne 
\sigma _2$. Also, if the unperturbed energies are chosen complex, either in
addition or instead of the choice just made, the qualitative picture remains.
The switching from level avoidance to level crossing of the real parts, --and 
associated with it the switching from crossing to avoidance of
the related imaginary parts--, is effected by the slipping of the EP over the
real $\lambda -$axis. Within the model considered here 
it can be achieved by starting 
with different unperturbed widths and/or an absorptive part of the Hamiltonian
which couples to the two channels with different strength. In Fig.2 it is
demonstrated for a four dimensional model, that our findings are not just the 
fluke of a two dimensional model \cite{cdemb}.

\begin{figure}
\epsfxsize=3.5in
\centerline{
\epsffile{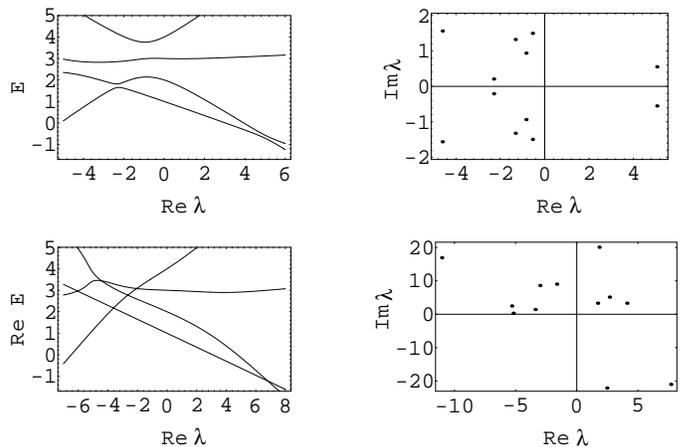}}
\vglue 0.15cm
\caption{
Effect of strong absorption in an arbitrary four level model. Complex 
conjugate pairs of EPs of a non-absorptive Hamiltonian (top left) lead to 
genuine level repulsion (top right). The absorption moves one member of each 
pair of EPs either into the  upper or lower $\lambda -$plane (bottom left) 
thereby effecting level crossing (bottom right).
}
\label{fig2}
\end{figure}

The top row shows a usual level repulsion 
among the four levels together with the exceptional points without absorption,
i.e.~for a real symmetric Hamiltonian,
while in the bottom row the absorption has been made sufficiently strong to
enforce the crossing of all levels; the last EP which has just slipped over the
real axis with increasing absorption lies at $\lambda \approx -5.5$. Note the 
symmetric positions of the EP
with respect to the real axis in the top row which prevails for a selfadjoint 
Hamiltonian \cite{ant}. Also note that the crossing of the real axis by an EP
can happen in either direction; this is why, in the particular case considered,
two EP are left in the lower half plane for the absorptive Hamiltonian.

We present a topological argument why either the real parts or the imaginary
parts must cross when energy trajectories of an absorptive
Hamiltonian pass the vicinity of an EP. We denote
by $\lambda _{\rm cross}$ the real part of an EP and follow
the trajectories $E_{1,2}(\lambda )$ for real values of 
$\lambda $ in the interval $[\lambda _{\rm cross}
-\delta ,\lambda _{\rm cross}+\delta ]$. In Fig.3 the complex numbers
$E_{1,2}(\lambda _{\rm cross}-\delta)$ are indicated by $A$ and $A'$. By
definition they have different real and imaginary parts. The endpoints of
the trajectories, which are at $E_{1,2}(\lambda _{\rm cross}+\delta)$ and
denoted by $B$ and $B'$, must schematically be situated as indicated, since
we consider the vicinity of a square root singularity. Schematically we may 
assume the singularity to be in the middle of the square like figure. When
$\lambda $ is sweeping over the interval  $[\lambda _{\rm cross}
-\delta ,\lambda _{\rm cross}+\delta ]$ the energy trajectory starting at $A$
can move to $B$ in which case $A'$ must move to $B'$ (dotted lines). 
This is the case of
width crossing and level avoidance. The other possibility is that the 
endpoints  are interchanged which is level crossing and width avoidance 
(dashed lines). Only
in the special case where the parameter $\lambda $ moves straight through
an EP will both, the real and imaginary part, cross. Since it is a square 
root singularity the angles between the in and out trajectories at the EP
must be at 90 degrees in the energy plane (solid lines).

\begin{figure}
\epsfxsize=3.0in
\centerline{
\epsffile{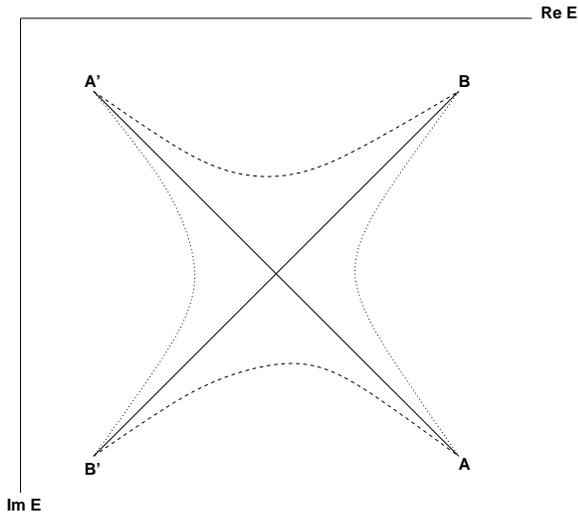}}
\vglue 0.15cm
\caption{
Schematic energy trajectories in the complex energy plane. Explanations in
main text.
}
\label{fig3}
\end{figure}

The different behaviour of the energy trajectories, depending on a left hand or
right hand side passage of an EP, is also reflected in a different behaviour of
the corresponding wave functions. It can be intuitively argued, and we confirm
this formally below, that we should expect the phase of {\sl one but not the
other} wave function to be different when comparing them behind the point of
anti-crossing and crossing of their real parts (Fig.1). 
In fact, for the situation of level avoidance, it is well
known that the wave functions after the point of repulsion are basically as
if the levels would have crossed but for a minus sign of one. In the vicinity
of the point of repulsion the eigenvectors can be parametrised by an angle
$\alpha $ which ranges from $0$ to $\pi /2$ when $\lambda $ is sweeping over
the repulsion point. Denoting by $\psi _1$ and $\psi _2$ the wave functions
of the top and bottom level before the repulsion, respectively, we have
\begin{eqnarray} \nonumber 
\psi _{\rm top}=\psi _1\cos \alpha -\psi _2\sin \alpha \\
\psi _{\rm bot}=\psi _1\sin \alpha +\psi _2\cos \alpha 
\end{eqnarray}
for the wave functions behind the repulsion. In turn, if the
levels do cross, the wave functions pertaining to the same levels will not 
undergo this rotation. As a consequence, behind the crossing or anti-crossing 
the bottom level is associated with $\psi _1$ while the wave function of 
the top level is $+ \psi _2$ for the crossing and $-\psi _2$ for the 
anti-crossing. This intuitive argument does not really specify which of the 
two wave functions changes its sign; what matters is the relative
change of one of the wave functions. Which one it is in the final comparison
depends on the sense of direction by which the EP is encircled.

To confirm more formally the statement about the phases we consider the two
situations displayed in Fig.1. We parametrise the
state vectors by the complex angle $\theta $, {\it viz.}
\begin{equation} \label{psi}
\psi _1(\lambda )=\pmatrix{\cos \theta \cr \sin \theta }, \quad
\psi _2(\lambda )=\pmatrix{-\sin \theta \cr \cos \theta }.
\end{equation}
with  
\begin{eqnarray} \label{tan}
\tan \theta(\lambda )&=&   \\ \nonumber
(&\lambda & (\omega _1-\omega _2)\sin 2 \phi _1 
-i\mu (\sigma _1-\sigma _2)\sin 2 \phi _2)/  \\ \nonumber
&(&E_1(\lambda )  -E_2(\lambda )+\epsilon _1-\epsilon _2+  \\ \nonumber
&\lambda &(\omega _1-\omega _2)\cos 2 \phi _1
-i\mu (\sigma _1-\sigma _2)\cos 2 \phi _2).
\end{eqnarray}
For the value of $\mu $ (and $\phi _2=0$) which yields the anti-crossing 
we read off from Eq.(\ref{tan})
the expected result: $\theta (0)=0$ and $\theta (\lambda )\to
\phi _1$ for $\lambda \gg |(\epsilon _1-\epsilon _2)/(\omega _1-\omega _2)|$.
In obtaining this result use is made of
$E_1-E_2=2R\to \lambda (\omega _1-\omega _2)$
for $\lambda \gg |(\epsilon _1-\epsilon _2)/(\omega _1-\omega _2)|$. 
For the other value of $\mu $ yielding the crossing of the levels we
now have to observe that we crossed into the other sheet of the square root
which means 
$E_1-E_2=-2R\to -\lambda (\omega _1-\omega _2)$. As a consequence we find
this time $\tan \theta \to \tan (\phi _1+\pi/2 )$ which confirms the
result. This consideration also clarifies that it is the square root
singularity that brings about this particular phase change. The values of
$\lambda $ which exceed $\lambda _{\rm cross}$ get us into different Riemann
sheets depending on whether we pass the EP on its right hand or left hand side.

The different cases as illustrated in Fig.1 have already been
experimentally established in an electro-magnetic resonator \cite{phil}. Two 
coupled resonators have been used as experimental setup. The levels of the one 
have been tuned by a parameter which plays the role of our parameter 
$\lambda $. As a second parameter, the coupling strength between the 
resonators has been controllable; we
denote this quantity by $x$. For fixed absorption which is achieved
by suitable antennas the necessary different widths have been adjusted. The
situation has also been modelled by two levels, however without reference 
to the existence of EPs. With the parametrisation used 
in the present paper it corresponds to the choice $\mu=0, \phi _1=\pi /4$
and $\epsilon _k=E_k-i\Gamma _k/2$ with $\Delta \Gamma =\Gamma _1- \Gamma _2
\ne 0$. The exceptional points are then situated at
$$ \lambda _c=-\Delta E-i\Delta \Gamma \pm i x$$
with $\Delta E=\epsilon_1 -\epsilon _2$. In this way, the difference 
$\Delta \Gamma $ and/or the coupling $x$ can be
adjusted such that one EP lies just above or below the real $\lambda -$axis 
thus giving rise to the two cases illustrated in Fig.1. The equipment
used in \cite{phil} did not allow a measurement of the phases of 
the wave functions,
i.e.~of the field strengths. It appears, however, that this is  possible
\cite{arich}.

We stress that exceptional points are a universal phenomenon in contrast to
diabolic points. While diabolic points may arise when two real parameters
are suitably chosen in a Hamiltonian, exceptional points always occur 
whenever there is
level repulsion. The physically interesting aspect is of course the access
to one or more of these points in an experiment. It was demonstrated that this 
is achievable in dissipative resonators. It is expected that it should be
possible in a variety of other systems, systems where interacting resonances
prevail. The present paper focusses on level crossing or anticrossing and
related phase behaviour for the associated wave functions. Interference
effects between the two distinct cases -- left hand and right hand passage 
of an EP -- and statistical aspects
for large number of resonance states will be the subject of forthcoming
considerations.

{\bf Acknowledgment}
The author acknowledges useful discussions with Peter von Brentano as well
as the warm hospitality in the Theory Group of the Max Planck Institute at
Heidelberg where most part of this paper was written.

\end{document}